# An Automated Home Made Low Cost Vibrating Sample Magnetometer

S. Kundu and T. K. Nath[*]

*Department of Physics and Meteorology, Indian Institute of Technology Kharagpur, Kharagpur-721302*
[*]*email: tnath@phy.iitkgp.ernet.in*

**Abstract.** The design and operation of a homemade low cost vibrating sample magnetometer is described here. The sensitivity of this instrument is better than $10^{-2}$ emu and found to be very efficient for the measurement of magnetization of most of the ferromagnetic and other magnetic materials as a function of temperature down to 77 K and magnetic field upto 800 Oe. Both M(H) and M(T) data acquisition are fully automated employing computer and Labview software.

**Keywords:** Vibrating sample magnetometer, Magnetization.
**PACS:** 07.55.Jg, 75.60.Ej

## INTRODUCTION

The vibrating sample magnetometer is the most widely used instrument for the measurement of magnetization. After the invention of this technique by S. Foner [1] there have been many modification and improvement of this instrument [2]. However, the basic principal remains the same and is very simple. In this technique, the magnetic specimen is vibrated with a certain frequency and amplitude employing a vibrator. At the same time a static magnetic field is applied to magnetize the sample. Now, a set of pick up coils are kept near the sample to pick up the signal voltage induced in the coils due to the change in the magnetic flux linked, caused by the vibration of the magnetized sample. Now - a - days the phase sensitive detection technique is used to detect the induced signal employing commercial Lock-in-amplifiers.

## DESIGN

Design of this magnetometer is kept as simple as possible by means of using commonly available materials. Schematic of the set up is shown in the Fig. 1. A brass plate is used as the main platform of the set up. A loud speaker is kept on this brass platform and used as the vibrator. A long glass tube, protected by a stainless steel cover is attached below this brass plate and kept fastened down to the liquid nitrogen container. The pickup coil assembly is fixed at the end of this glass tube. A solenoid is also placed

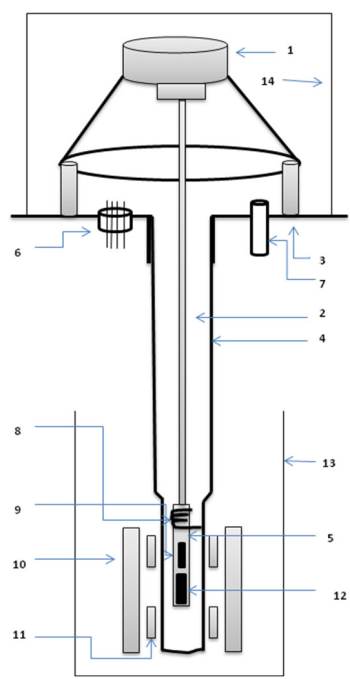

**FIGURE 1.** Schematic of the VSM (1) loud speaker, (2) sample holder rod, (3) brass plate, (4) glass tube, (5) sapphire sample holder, (6) electrical feed through, (7) vacuum port, (8) heater wire, (9) cernox temperature sensor, (10) primary coil, (11) pickup coils, (12) sample, (13) liquid nitrogen container, (14) perspex cover.

covering the pair of pickup coils coaxially to generate the static magnetic field.

Each of the pickup coils is having approximately 3000 turns of Cu wire and separated by a distance of 16 mm. Width of each of the coils is 6 mm. A long thin stainless steel tube is used as sample holder rod which is attached to the vibrator at one end. A sapphire plate is fitted to the other end of the rod to serve as the sample holder. A non-magnetic heater and cernox thermometer are placed on the sapphire plate to control the sample temperature. Both the coil systems (solenoid and pickup) are dipped in the liquid nitrogen during measurements. A SR830 Lock-in-amplifier (LIA) is used to measure the induced signal in the pickup coils locked at the same driving frequency.

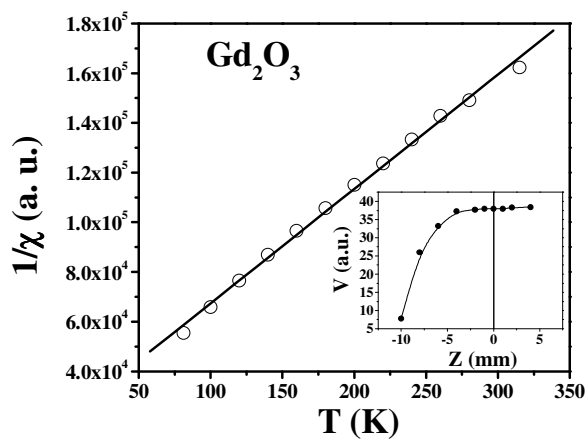

**FIGURE 2.** Plot of inverse susceptibility ($\chi$) of $Gd_2O_3$ as a function of temperature. Inset shows the signal induced in the pickup coils as a function of distance along the axis showing the constant signal zone.

## OPERATION AND CALIBRATION

The Sine Out of the LIA is used as the necessary electrical signal for the vibration of the loud speaker. As the magnetized sample placed on the sample holder is vibrated a signal is induced in the pickup coils. This signal has the same frequency as that of the vibration. The LIA is used to precisely measure this signal which is proportional to the magnetic moment of the sample. DC field in the solenoid is generated by employing a Keithley source meter unit SMU-2612. All the electronic equipments are interfaced to the computer using a data acquisition card and Labview software.

One important issue of such design is getting a constant signal zone between the pair of pickup coils. To map the induced signal we have measured it as a function of the distance along the coil axis (Z). Z=0 corresponds to the centre of the gap between the coils. A constant signal zone between the coils shows the perfection of our design (Inset of Fig. 2). This also allows us to put the sample at Z=0. Linearity of the set up was checked by measuring the magnetic moment of high-moment paramagnetic salt $Gd_2O_3$. Plot of inverse susceptibility with respect to temperature is linear as depicted in the Fig. 2 (following Curie law). This indicates the excellent quality of the set up which is free from any significant noise coupling and temperature error. The calibration of the set up was done conventionally by employing pure Ni sample.

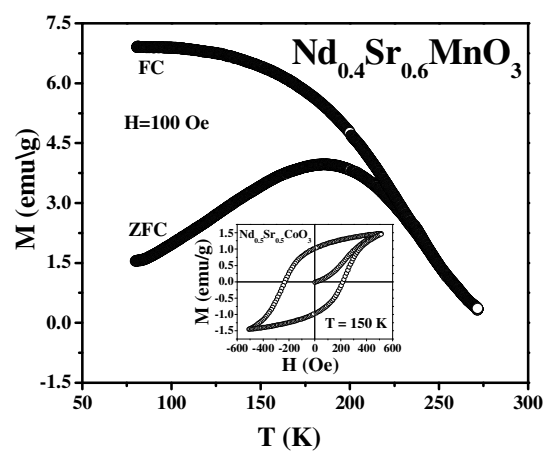

**FIGURE 3.** Plot of M (H) and M (T) measurements carried out in the magnetometer. Main panel shows the zero field cooled (ZFC) and field cooled (FC) magnetization of nanoparticles of $Nd_{0.4}Sr_{0.6}MnO_3$. Inset shows the M-H loop of $Nd_{0.5}Sr_{0.5}CoO_3$.

Some typical measurements carried out employing this set up are shown in the Fig. 3. The main panel shows the temperature dependent magnetization of $Nd_{0.4}Sr_{0.6}MnO_3$ whereas the inset shows a hysteresis behavior of $Nd_{0.5}Sr_{0.5}CoO_3$. Both the data clearly demonstrate that they are of very high quality and evidently free from fluctuations. Overall sensitivity of the set up was found to be better than $10^{-2}$ emu. Temperature stability of the set up was observed to be better than 50 mK.

## REFERENCES

1. S. Foner, *Rev. Sci. Instr.* **30**, 548-558 (1959).
2. R. V. Krishnan and A. Banerjee, *Rev. Sci. Instr.* **70**, 85-91 (1999).